\documentclass{article}

\usepackage{arxiv}

\usepackage[utf8]{inputenc} 
\usepackage[T1]{fontenc}    
\usepackage{hyperref}       
\usepackage{url}            
\usepackage{booktabs}       
\usepackage{amsfonts}       
\usepackage{nicefrac}       
\usepackage{microtype}      
\usepackage{lipsum}		
\usepackage{graphicx}
\usepackage{natbib}
\usepackage{doi}

\usepackage{amsmath}
\usepackage{float} 
\usepackage{enumerate}
\usepackage{multirow}

\title{Land use identification through social network interaction}

\author{ \href{https://orcid.org/0000-0002-3549-0279}{\includegraphics[scale=0.06]{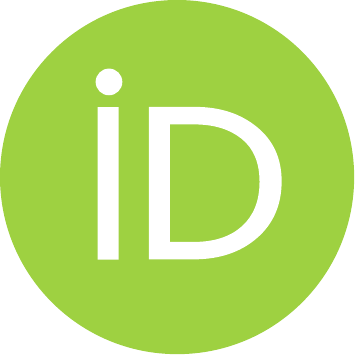}\hspace{1mm}Diana C. Pauca--Quispe}\\
		Department of System Engineering and Informatics\\
		Universidad Nacional de San Agustín de Arequipa\\ 
		Arequipa, Perú\\
	\texttt{dpaucaqu@unsa.edu.pe} \\
	\And
	\href{https://orcid.org/0000-0002-5485-9772}{\includegraphics[scale=0.06]{orcid.pdf}\hspace{1mm}Cinthya Butrón--Revilla} \\
		Department of System Engineering and Informatics\\
		Universidad Nacional de San Agustín de Arequipa\\ 
		Arequipa, Perú\\
	\texttt{cbutronr@@unsa.edu.pe} \\
	\And
	\href{https://orcid.org/0000-0001-6654-5976}{\includegraphics[scale=0.06]{orcid.pdf}\hspace{1mm}Ernesto Suarez--López} \\
		Department of System Engineering and Informatics\\
		Universidad Nacional de San Agustín de Arequipa\\ 
		Arequipa, Perú\\
	\texttt{esuarezlo@@unsa.edu.pe} \\	
	\And
	\href{https://orcid.org/0000-0002-1438-2615}{\includegraphics[scale=0.06]{orcid.pdf}\hspace{1mm}Karla Aranibar--Tila} \\
		Department of System Engineering and Informatics\\
		Universidad Nacional de San Agustín de Arequipa\\ 
		Arequipa, Perú\\
	\texttt{karanibar@unsa.edu.pe} \\	
	\And
	\href{https://orcid.org/0000-0002-2666-293X}{\includegraphics[scale=0.06]{orcid.pdf}\hspace{1mm}Jesús S. Aguilar--Ruiz} \\
		Universidad Pablo de Olavide\\
		ES-41013, Seville, Spain\\
	\texttt{aguilar@upo.es} \\	
}

\hypersetup{
pdftitle={Land use identification through social network interaction},
pdfauthor={Diana C. Pauca--Quispe, Cinthya Butrón--Revilla, Ernesto Suarez--López, Karla Aranibar--Tila, Jesús S. Aguilar--Ruiz},
pdfkeywords={Land use, Social networks data, Natural Language Processing, Classification},
}

\begin{document}
\maketitle

\begin{abstract}
The Internet generates large volumes of data at a high rate, in particular, posts on social networks. Although social network data has numerous semantic adulterations, and is not intended to be a source of geo--spatial information, in the text of posts we find pieces of important information about how people relate to their environment, which can be used to identify interesting aspects of how human beings interact with portions of land based on their activities. This research proposes a methodology for the identification of land uses using Natural Language Processing (NLP) from the contents of the popular social network Twitter. It will be approached by identifying keywords with linguistic patterns from the text, and the geographical coordinates associated with the publication. Context--specific innovations are introduced to deal with data across South America and, in particular, in the city of Arequipa, Peru. The objective is to identify the five main land uses: residential, commercial, institutional--governmental, industrial--offices and unbuilt land. Within the framework of urban planning and sustainable urban management, the methodology contributes to the optimization of the identification techniques applied for the updating of land use cadastres, since the results achieved an accuracy of about 90\%, which motivates its application in the real context. In addition, it would allow the identification of land use categories at a more detailed level, in situations such as a complex/mixed distribution building based on the amount of data collected. Finally, the methodology makes land use information available in a more up-to-date fashion and, above all, avoids the high economic cost of the non--automatic production of land use maps for cities, mostly in developing countries.
\end{abstract}


\section{Introduction}
The current dynamic nature of cities has generated substantial changes in urban environments, making the analysis of geo--spatial information more complex \citep{Liu&Xi}. The urban management and  planning of cities has been nourished for years by Geographic Information Systems (GIS) tools, since they allow storing, modeling, and analyzing geo--spatial data. On the other hand, urban planning faces challenges such as organizational changes, staff availability, updating resources and data availability \citep{yeh1999urban}. In this context, land use maps are one of the most demanded resources by authorities and researchers, for conducting urban planning and environmental sustainability studies \citep{anugraha_land_2018}. However, the production of these maps is costly and very time--consuming \citep{Ye}. Consequently, the production and management of land use maps evidence a technical problem \citep{sendra_uso_2000} cau\-sed by economic constraints, mostly in developing countries.

The Internet generates large volumes of data at a high rate, in particular, posts on social networks. Social networks have become a source of data for researchers, because this data contains information that allows studying the interaction of citizens with their environment \citep{geertman_can_2019}. The concept of ``user--generated content'' \citep{thakur,lei_spatial-temporal_2018}, which encompasses how citizens share excerpts from their daily lives, can be translated to different application domains: the representation of urban boundaries, relationships between weather conditions and traffic, citizen's activity patterns, urban transportation behavior and land use \citep{mora_analysis_2018,geertman_can_2019}, urban planning and spatial planning. Therefore, social networks are a valuable source of data that could be used to identify interesting aspects of a city's land use. 

Although social network data has numerous semantic a\-dulterations, and is not intended to be a source of geo--spatial information, in the text of posts we find pieces of important information about how people interact with their space, and can complement other sources in the process of identifying land uses \citep{tessore}. These data can be composed of various fields such as message text, location, and images. In this context, various approaches have been developed and applied to analyze this data type, being semantic analysis of the text of the messages  --unstructured information--  the basis of much of the research related to social networks. However, it is an underexploited resource in obtaining and analyzing geographic information \citep{stock_mining_2018}. Current land use identification and analysis studies rely primarily on publication metadata, such as time and geographical coordinates --structured information--.

The various studies on land use classification have rarely considered the content of posts as a source of data to infer the type of use. In the content of social network posts, people communicate attitudes, activities, relationships and other emotional states that are often complemented with a geo--spatial location. In written expressions it is possible to find words that allow inferring information about where a user is situated. For example, if a person posts the following message ``having lunch with friends'' on a social network, we could assume that they are in a restaurant, a house, a shopping mall, or another place where they can have lunch. In addition, if we associate two or more words such as "having lunch with friends in a restaurant", the connector ``in'' and the word ``restaurant'' provide us with a high probability that the place where he/she is located is, in fact, a restaurant. If the coordinates associated with the message are included, we can know the space where the restaurant is placed, and consequently, the type of land use to which that space belongs. Therefore, the potential of the content of social network publications to address the study of land use is remarkable. 

Nevertheless, the language used in social network publications is very informal, includes many idioms, expressions specific to each region, misspellings, or other language deformations. For this reason, the processing of this type of information with specific objectives becomes an important challenge \citep{Iglesias}, in which numerous algorithmic techniques will concur, to form a methodology capable of extracting the valuable knowledge.

Most studies on land use are mainly oriented to classification tasks. When we use a data source such as the expressions of a language, it is necessary to propose a methodology that starts with data collection, followed by pre--processing and, finally, classification. The fundamental reason is that the expressions in each language and country --and even re\-gion-- are different. For example, in order to carry out the present study, it was necessary to create a corpus and a dictionary for the Spanish language and for expressions specific to Peru.

This paper proposes a methodology for the identification of land uses using Natural Language Processing (NLP) from the contents of the popular social network Twitter. It will be approached by identifying keywords with linguistic patterns from the text, and the geographical coordinates associated with the publication. The methodology is composed of the following stages: data collection, corpus and dictionary creation, pre--processing, feature extraction, learning, prediction and representation. At each stage, context--specific innovations are introduced to deal with data from South America and, in particular, the city of Arequipa, Peru. The objective is to identify the five main land uses: residential, commercial, institutional--governmental, industrial--offices and unbuilt land.

Within the framework of urban planning and sustainable urban management, the methodology proposed in this study contributes to the optimization of the identification techniques applied for the updating of land use cadastres, since the results achieved an accuracy of about 90\%, which motivates its application in the real context. In addition, this model would allow the identification of land use categories at a more detailed level, in situations such as a complex/mixed distribution building based on the amount of data collected. Finally, the methodology makes land use information available in a more up-to-date and, above all, much less costly way. 

The document is organized as follows: first, a review of the recent scientific literature related to the use of machine learning techniques applied to the categorization of land use, and specifically involving data from social networks, followed by an overview of the NLP--based methodology; next, a description of the application context in the city of Arequipa, containing examples of the results provided by the main phases of the methodology, together with the results achieved in comparison to several models; also, visual examples are presented to illustrate the contrast between results from our approach and what is registered in the cadastre; finally, conclusions and future work.

\section{State--of--the--art}

\subsection{Land use and urban planning applications}

Recent proposals for urban growth cover aspects such as the potential exploitation of land uses: those that allow to accommodate a high urban diversity. Land uses are thus defined as a measure of the diversity of uses contained in a given space \citep{hajna_call_2014}, which allows in a specific way to identify nearby uses or nearby activities developed in a limited spatial range. The term \textit{land use} refers to the employment that human beings give to a portion of land based on the activities carried out in it. However, in order to use land uses in urban management and planning, an updated map with the most recent and accurate information is necessary \citep{Ye,Terroso2020} to make the right decisions \citep{DaSilva}. Under this situation, several terms emerge, such as \textit{urban computing}  \citep{Silva} and urban planning applications \citep{Frias}, which have become an emerging research area where urban problems in cities are studied using different data sources such as electronic devices -mainly smartphones-, location--based social media, web pages -e.g., analyzed with Robotic Process Automation technologies-, among other digital information.

Many approaches include the use of images (e.g. from satellites), sometimes together with some structured data (e.g. points of interest). For example, Liu et al. \citep{Liu&Xi} use natural physical features from high spatial resolution images (HSR) and socio--economic semantic features (frequent characteristic words related to an urban land type) from social data to create a dictionary of land use words, including data from multiple sources such as OpenStreetMap road networks, Gaode's Points of Interest and Tencent's real--time user density. On the other hand, Zhan et al. \citep{zhan_inferring_2014} use large--scale Twitter log data, and they propose a preprocessing method in which the raw data only contains coordinates and activity category information. One of the few papers that mentions data preprocessing is \cite{thakur}, who used Twitter posts and metadata. From each tweet they use text to analyze whether a space is a restaurant, airport, or stadium. They perform preprocessing of the text of each tweet before using a term frequency--based technique, previously removing emoticons and other non--ASCII characters, as well as hashtags and the ``@'' character.

\subsection{Social Networks and Text Mining}

The incremental use of social networks has been producing large amounts of data that are being extracted, analyzed, and structured \citep{stock_mining_2018}. This allows gathering useful information that can reflect different areas such as the human dynamics in a city, identifying how people live and interact with the environment, health applications, natural hazards management, tourism, environmental monitoring, crimes and disturbances \citep {garcia_palomares}.

The ways to extract the information offered by a publication on social networks range from the analysis of its metadata (geo--tags, time and date, username, place name, etc.), the profile of the user who publishes it, to the application of text mining to the message of the publication \citep{Iglesias}. The message is considered the backbone of many of the investigations related to the inference of location because of its enormous potential \citep{ajao}, apart from presenting several scientific challenges. Since the texts of social media are mostly generated by mobile devices and have no writing restrictions, it gives the user a great margin of typographical error and brevity. These texts also include links, emoticons, use of informal language and idioms, spelling and grammatical errors, presence of user mentions and hashtags, HTML tags, use of acronyms and abbreviations. Therefore, the main challenge is to clean up the large amount of noise in the messages, in addition to deal with the unstructured format, in contrast with articles, web pages, or blogs, that have more content and make use of conventional grammar and semantic rules.

\subsection{Text Mining}

Text mining is a research field that aims to automatically discover or extract new knowledge from texts written in natural language \citep{yuyun,tandel}. According to the objective to be achieved, the text mining process has a set of stages that are classified into data collection, pre--processing, transformation, and analysis.

\subsubsection{Data collection}

The collection is the first step of the text mining process, where unstructured data is captured from different sources such as blogs, reviews, news, publications on social media, among others. The data is stored for future pre--processing and analysis.

\subsubsection{Pre--Processing}

The objective is to obtain clean and actionable data from the data collected through the application of different debugging and cleaning techniques. There is no specific order to perform the pre--processing task, so this must be determined empirically \citep{rexiline}, experimenting with each technique individually, comparing the results, and combining those that perform best. The pre--processing of social media data presents certain challenges due to the use of informal language and the length of the messages (for instance, tweets are very short in length) \citep{Kateb}.

The techniques commonly used in text pre--processing are the following \citep{lansley, tellez, varma, stock_mining_2018, tessore, Kulkarni}:

\begin{enumerate}[a)]
\item Normalization and noise reduction: this process unifies the text and removes irrelevant and meaningless elements. The literature recommends the deletion of URLs, HTML tags, special characters, emojis, or emoticons.
\item Remove punctuation marks: as punctuation marks do not add additional information, the elimination of them helps reduce the size of the data and increase the model efficiency.
\item Remove stopwords: these are terms that help build sentences but do not commonly provide meaning, like prepositions or articles.
\item Processing of abbreviations, acronyms, and entities.
\item Spell correction.
\item Tokenization: it splits text into significant words delimited by blank spaces, commas, periods, or any other special character.
\item Lemmatization: it is the search for the words' lemma to unify the terms that give the same information (for example, derivational forms of verbs).
\item Part of Speech Tagging (PoS): grammatical tagging of each word.
\end{enumerate}

\subsubsection{Data transformation}

Data transformation involves the extraction and selection of characteristics (SC) from text data (string) to generate a suitable representation for computational learning. In classification, the transformation is crucial as the SC directly impacts the result of the classifier. The selection depends on the type of document and the classification chosen. The objective is to find characteristics with relevant information that improve the precision of the classifier. Then, the text is represented as a value using a binary representation or as a SC technique, such as Term Frequency (TF), Inverse Document Frequency (IDF), or Term Frequency--Inverse Document Frequency (TF--IDF). From these representations, the global feature space is generated from all the texts used for training. The next paragraph pre\-sents some techniques to obtain the characteristics of a text:

\begin{enumerate}[a)]
\item Term Frequency--Inverse Document Frequency (TF--IDF): it seeks to find the balance between the Frequency Term (importance of a word $w_{k}$ locally in a text $T$), and IDF (a global measure of the importance of $w_{k}$ in the corpus) \citep{tellez}.
\item N--grams: they are sequences of words grouped according to the value given to $N$. For example, the 1--grams (unigrams) of $T$ = \textit{``Tomorrow is Thursday''} are  $W_{1}^{T}=\{$\textit{Tomorrow, is, Thursday}$\}$, the 2--grams (bigrams) are $W_{2}^{T}$ = $\{$\textit{Tomorrow is, is Thursday}$\}$; then, given a text $T$ with $m$ words, a set of n--grams of size $m-n+1$ is obtained.
\item Bag--of--PoS: application of n--grams in the PoS labels of the text to be analyzed.
\end{enumerate}

\subsubsection{Data analysis}

There exist different methods to analyze unstructured da\-ta, which are classified into information extraction, summary, grouping, and categorization \citep{maheswari,tandel}. The present study uses categorization in order to classify a set of documents into topics or categories, which requieres a correct combination of NLP and machine learning techniques.

In the field of machine learning, classification tasks are mainly supervised, which consist of training the system and then testing with information about the classes before the real classification process \citep{thangaraj}. Among the most popular techniques are Logistic Regression, Naive Bayes, Support Vector Machines, Decision Trees, Neural Networks, or k--Nearest Neighbor. The Naive Bayes approach is one of the most employed classifiers for analyzing text documents. It works under the principle of using the probabilities of words and categories that allow determining the classes for given documents, and assumes two interesting properties that makes the technique very fast: conditional independence among variables with respect to the target variable, and normal distribution of independent variables. In summary, Bayesian classifiers are simple and powerful in terms of the degree of certainty, which makes them a god choice for approaching NLP problems.

%
%
%
%

\section{Materials and Methods}

\begin{figure}[t]
	\centering
    	\includegraphics[width=1\linewidth]{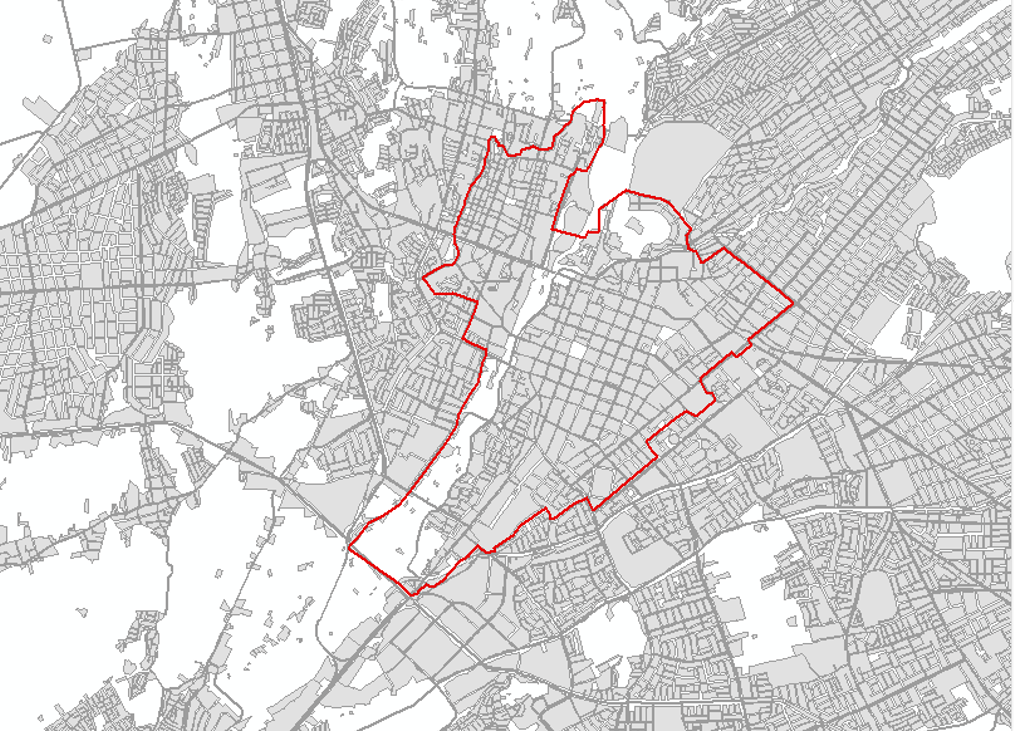}
	\caption{Map of the historical center of the city of Arequipa (own elaboration from data based on PlaMCha, 2017--2027).}
	\label{FIG:1}
\end{figure}

The area of study is the city of Arequipa, which is the second most populated urban area in Peru, with more than a million of inhabitants and a yearly growth rate of 2.3\%, according to the last census conducted in 2017. This study covers the historical centre of Arequipa (see Fig. \ref{FIG:1}), with an approximate area of 3.46 square kilometers and about 2251 properties located within 56 blocks. The categorization made is based on the real estate properties declared in the area.

The Master Plan for Arequipa's Historic Center and Surrounding Area (PlaMCha 2017--2027) describes the geogra\-phical characteristics of the historical center, which has three defined zones; old zone, monumental zone, and buffer zone located in the coordinates S: 16$^{\circ}$  23' 53.33" W: 71$^{\circ}$  32' 12.67" \citep{Anglada2018}. The oldest area, where the main monuments and the Main Square of the city are located, is called Damero (transl. checkboard) and is constituted by blocks of 111.4 \textit{m} per side and separated by streets of 10.3 \textit{m}, which gives the characteristic unitary image and covers an area of \begin{math} 1.41 \hspace{0.1cm}km^{2} \end{math} (141 ha.). The Monumental Zone has \begin{math}2.12 \hspace{0.1cm}km^{2}\end{math} (212 ha.), and the Buffer Zone has \begin{math}3.46 \hspace{0.1cm} km^{2}\end{math} (346 ha.) and covers the first 2 zones.  

The historical center was selected because it brings together different types of land use such as historical, artistic, cultural, workplaces, residential, commercial, and public spaces. The large influx of tourists and residents in this area makes it possible to collect a large amount of data from social networks, unlike in other parts of the city.

\subsection{Data Sources}

\subsubsection{Types of land use}

The land use map of the historical center of Arequipa, made by the PlaMCha, defines 14 categories of land use (see Table \ref{tbl1} -- left). However in this work, the re--categorization carried out by \citep{hajna_call_2014} is considered, which initially takes into account 48 categories of land use and then groups them into 5: residential, commercial, industrial--offi\-ces, institutional--government and unbuilt land (see Table \ref{tbl1} -- right). The residential category groups houses, apartments, and condominiums. The commercial category comprises properties that carry out typical activities of stores, bookstores, shopping centers, restaurants, bars, hotels, lodgings, parking lots, and entertainment venues. The industrial--offices category groups industrial, companies and offices land uses. The institutional--government category includes buildings related to education (institutes, academies, schools and universities), health (hospitals and clinics), cultural (cultural centers and museums), management (administrative, financial or governmental grounds) and religious centers. Finally, the category of unbuilt land is mostly composed of agricultural land (crops), vacant land (buildings that do not exceed 1\% of the surface), the Chili river course -which runs north--southwest through the historical area-.

\begin{table}[t]
\caption{Re--categorization of land use categories \citep{hajna_call_2014}.}\label{tbl1}
\centering
\begin{tabular*}{4in}{l@{\extracolsep{\fill}}l}
\toprule
Categories & Re--categorization \\
\midrule \midrule
Residential & Residential \\
\hline
Commerce & \multirow{3}{*}{Commercial} \\
Lodging \\
Parking\\
\hline
Industry & Industrial -- Offices \\
\hline
Education & \multirow{5}{*}{Institutional -- Governmental} \\
Health \\
Cultural  \\
Administrative  \\
Religious  \\
\hline
Vacant land & \multirow{4}{*}{Unbuilt land } \\
Crop\\
Chili River  \\
Others &  \\
\bottomrule
\end{tabular*}
\end{table}

\subsubsection{Cartographic map}

This study focuses on determining the categories of land use in the historical center of Arequipa, hence it is necessary to define the boundaries and characteristics of the area as a function of polygons.The information is obtained from the documents developed in the PlaMCha, and the cadastral maps of the historical center developed by the Technical Team of the Municipal Planning Institute in the Pilot Project named ``Altura para la Cultura'' (transl. Height for Culture) \citep{Anglada2018}. Cadastral maps were provided as GIS data, representing set of polygons for blocks and lots in the historical center (see Fig. \ref{FIG:1}).

\subsubsection{Geo--tagged Twitter publications}

Tweets allow the categorization of land uses where tweets were generated. A tweet is a short message of 280 characters maximum composed of text, emojis and attachments, published on the Twitter platform, which is considered one of the largest sources of information fed by millions of users \citep{emergency_2019}. Twitter Application Programming Interfaces (APIs) allow capturing tweets encoded in Ja\-vaScript Object Notation (JSON) format with their associated attributes and values. A tweet can have around 150 associated attributes, although this research only requires the ID, user ID, text, timestamp, geodata, and language of the message.

\subsection{Experimentation}

\begin{figure*}
	\centering
		\includegraphics[width=1\linewidth]{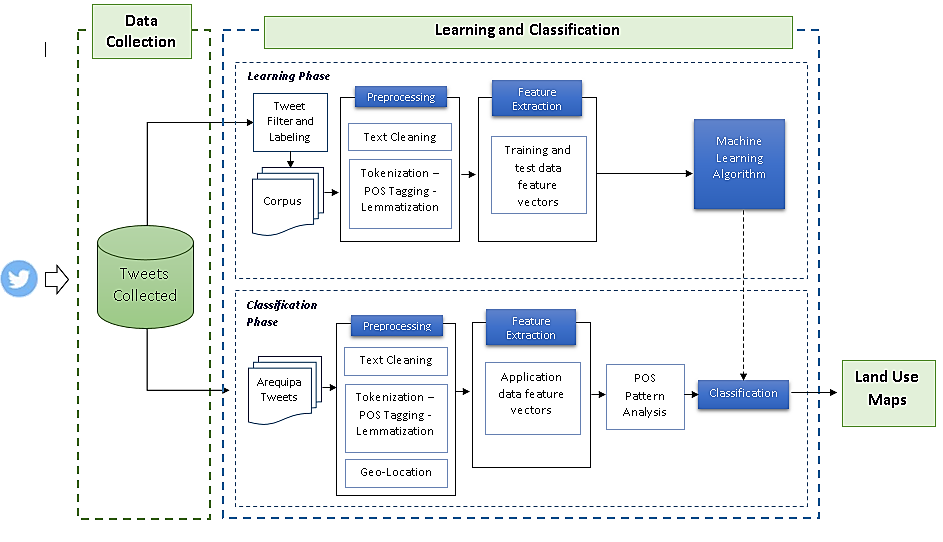}
	\caption{Outline of the proposed methodology.}
	\label{FIG:2}
\end{figure*}

The methodology is divided into a sequence of tasks, which begin with the data collection, followed by splitting the data for building the corpus and for validating. The validation data go through a previous classification using PoS patterns before entering the classification algorithm. For clearer interpretation, the results of the tagged data are presented on a map of the study area according to the related coordinates of the tweet. Finally,  quality metrics are shown to validate the approach. The methodology is graphically illustrated in Fig.\ref{FIG:2}. 

\subsubsection{Data Collection}

In this phase, the model connects to the Twitter Streaming API using the Tweepy library to capture geo--referenced tweets across South America. The Streaming API has certain limitations, but it allows the download of 100\% of tweets that meet  the defined filter, as long as they are less than 1\% of the global volume of publications at a given time \citep{campan}. The application was run on a server for a period of 10 months (from April 2019 to February 2020 -up to the beginning of the pandemic-) and data was stored in a MySQL database. 

A corpus related to land use categorization was not available in Spanish, so it was necessary to build one. The collected data was divided into two sets: the first set consists of 3870 tweets located in the historical center of Arequipa, that will then be used to identify the type of land use. The second set is used for the creation of the corpus and consists of 42318 tweets located anywhere in South America.

These tweets are previously processed to remove the duplicates, blanks, single word and only numbers. 

As a result, we obtained a total of 24995 messages that were semi--automatically categorized into a land use type according to their content and geographical coordinates, leaving a total of 4538 useful tweets in different languages, which were randomly split into training and test. The result was distributed into the defined categories and divided into subcategories to avoid data imbalance (see Table \ref{tbl2}).

\begin{table*}[h]
\caption{Distribution of classes in the corpus.}
\label{tbl2}
\begin{tabular}{p{1.4cm}p{1.6cm}p{1.4cm}p{1.5cm}p{1.5cm}p{1.5cm}p{1.2cm}p{1.4cm}p{1.2cm}}
\toprule
\multicolumn{3}{c}{Commercial} & \multicolumn{3}{c}{Institutional -- Government} & \centering Industrial Offices & Residential & \centering Unbuilt land\tabularnewline
\cmidrule(lr){1-3}\cmidrule(lr){4-6}\cmidrule(lr){7-7}\cmidrule(lr){8-8}\cmidrule(lr){9-9}
\multicolumn{3}{c}{\textbf{2539}} & \multicolumn{3}{c}{\textbf{1212}} & \centering \textbf{138}& \centering \textbf{219} & \centering \textbf{431}\tabularnewline
\cmidrule(lr){1-3}\cmidrule(lr){4-6}\cmidrule(lr){7-7}\cmidrule(lr){8-8}\cmidrule(lr){9-9}
Commercial & \centering Commercial Restaurant & \centering Commercial Service & Institutional & \centering Institutional Education & \centering Institutional Cultural & \centering Industrial Offices & Residential & \centering Unbuilt land \tabularnewline
\cmidrule(lr){1-1}\cmidrule(lr){2-2}\cmidrule(lr){3-3}\cmidrule(lr){4-4}\cmidrule(lr){5-5}\cmidrule(lr){6-6}\cmidrule(lr){7-7}\cmidrule(lr){8-8}\cmidrule(lr){9-9}
\centering 1177 & \centering 874 & \centering 488 & \centering 497 & \centering 371 & \centering 344 & \centering 138 & \centering 219 & \centering 431 \tabularnewline
\bottomrule
\end{tabular}
\end{table*}

\subsection{Pre--Processing}


The pre--processing is divided into two phases: the first phase consists of cleaning the text to remove noise and correct data, and the second phase is the data filtering to identify only the tweets which are within a block of the historical center. The tweets to be used as classifier input go through the two pre--processing phases, while the corpus tweets only go through the first phase. 

\begin{table}
\centering
\caption{Text output after noise removal, translation, punctuation removal, and replacement of hashtags and mentions.}
\label{tbl3}
\begin{tabular}{p{7.8cm}p{7.8cm}}
\toprule
Original Text & Pre--Processed Text \tabularnewline
\midrule \midrule
I'm at Mallplaza Bellavista -  \text{@mallplazaperu} in Bellavista, Callao https://t.co/brtyxSe8CY	& \textbf{Estoy en} Mallplaza Bellavista \textbf{en mallplazaperu en} Bellavista Callao \tabularnewline
\hline
work breakfast \text{!} \text{\#friends}  \text{\#meeting} en Universidad Jorge Tadeo Lozano \text{https://t.co/sNYJhxG6cw}	& \textbf{Trabajo desayuno amigos reunión} en Universidad Jorge Tadeo Lozano \tabularnewline
\hline
Un dia cualquiera en Cevicheria Karloncho Oficia \text{https://t.co/f9kdEEwdMx} &	Un dia cualquiera en Cevicheria Karloncho Oficia \tabularnewline
\hline
He venido a que mami me atiborre de comidaaaaaa (\text{@Residencial} Parque Central in Lima) \text{https://t.co/dCCBbEgvZm} & He venido a que mami me atiborre de comidaaaaaa \textbf{en Residencial} Parque Central  \textbf{en} Lima \tabularnewline
\bottomrule
\end{tabular}
\end{table}

\begin{enumerate}[(a)]
    \item Noise removal, language detection, and Spanish translation. URLs, symbols, and HTML tags are removed \citep{salas_zarate, rexiline}. For translation and detection, the \textit{Googletrans} library for Python is used, which allows processing large amounts of records without a query limit; the translation is of the entire text and not word by word so that the result is a meaningful sentence, as Google Translate Ajax API does. Hashtags and mentions are excluded from the translation.
    \item Hashtag processing. Hashtags are words that tag a tweet to a topic that is generally related to its content \citep{asriadie}. In this study, these words are preserved and translated individually if they are in English.
    \item Elimination of punctuation marks and replacement of mentions. All punctuation marks are removed, except for the symbol \textit{@} which is replaced by the word \textit{in},which is useful in the context of location extraction from a text \citep{thakur}.
\end{enumerate}

An example of the result of applying the first three techniques is shown in Table \ref{tbl3}.

\begin{enumerate}[(d)]
    \item Processing of Abbreviations, Acronyms, Slang, and Establishment Names. The most recurrent abbreviations and slangs within the corpus are identified and stored for the construction of a dictionary. The names of establishments, institutions and acronyms identified from google maps are added to this list, which includes bars, pharmacies, hotels, universities, etc. located within the Historical Center of Arequipa. The resulting dictionary is used to identify these words in the text of the publication and replace them with the related word \citep{tellez}. The result is shown in Table \ref{tbl4}.

\end{enumerate}


\begin{table}[t]
\caption{Text output after processing abbreviations, acronyms, slang and establishment names.}
\label{tbl4}
\begin{tabular}{p{7.8cm}p{7.8cm}}
\toprule
Original Text & Abbreviation Processing \tabularnewline
\midrule \midrule
I'm at \text{Mallplaza} Bellavista - @\text{mallplazaperu} in Bellavista, Callao https://t.co/brtyxSe8CY & estoy en \textbf{centro comercial} bellavista en \textbf{centro comercial} en bellavista callao \tabularnewline
\hline
work breakfast! \#friends \#meeting en Universidad Jorge Tadeo Lozano https://t.co/sNYJhxG6cw	& trabajo desayuno \text{amigos reunión} en universidad jorge tadeo lozano \tabularnewline
\hline
Un dia cualquiera en \text{Cevicheria} Karloncho Oficia https://t.co/f9kdEEwdMx & un dia cualquiera en \textbf{restaurante} karloncho oficia \tabularnewline
\hline
He venido a que \text{mami} me atiborre de comidaaaaaa (@ Residencial Parque Central in Lima) https://t.co/dCCBbEgvZm & he venido a que \textbf{mamá} me atiborre de comidaaaaaa en residencial parque central en lima \tabularnewline
\bottomrule
\end{tabular}
\end{table}

\begin{enumerate}[(e)]
    \item Spell Checking. The open--source tool Hunspell is used (embedded in LibreOffice, Mozilla Firefox, Thun\-derbird and Google Chrome, and some proprietary software pac\-kages). It is a spell checker designed for languages with complex morphology, such as Spanish \citep{salas_zarate}. The concept of \textit{Longest Common Subsequence} (LCS) is applied to each of the options proposed by the spell checker to improve the result. Hunspell checks that each token in a publication is a valid word in the language; otherwise, it is replaced if the LCS value of the word proposed by the proofreader is not less than 71\%; if it is less, the word is deleted (see Table \ref{tbl5}).
\end{enumerate}

\begin{table}[t]
\caption{Spell--checking process.}
\label{tbl5}
\centering
\begin{tabular}{p{3.2cm}p{4cm}p{6cm}}
\toprule
Word to be corrected & Hunspell correction options & Choosing the best option using LCS \tabularnewline
\midrule
\multirow{6}{*}{sapato} & ['apasto', &  apasto = 50.0\%
\\ & 'zapato', & \textbf{zapato = 83.33\%}
\\ & 'patoso', & patoso = 66.66\%
\\ & 'topatopa', & topatopa = 66.66\%
\\ & 'sato', & sato = 50.0\%
\\ & 'pato'] & pato = 66.66\%
\tabularnewline
\hline
\multirow{3}{*}{ClubMilita} & ['Club Militar', & \textbf{Club Militar = 100.0\%} 
\\ & 'Club-militar', & Club-militar = 54.545\%
\\ & 'Militarizar'] & Militarizar =  63.636\%
\tabularnewline
\hline
\multirow{4}{*}{casiita} & ['casinita', & casinita = 57.142\%
\\ & 'casiterita', & casiterita =  57.142\%
\\ & 'marcasita', & marcasita =  57.142\%
\\ & 'canastita'] & canastita = 42.857\%
\tabularnewline
\hline
Munays & ['Ayunas'] & \text{Ayunas} = 50.0\% \tabularnewline
\bottomrule
\end{tabular}
\end{table}

\begin{enumerate}[(f)]
    \item Stopwords. All stopwords are removed, except the words ``en'' (transl. \textit{in}) and ``de'' (transl. \textit{from}) because these are spatial indicators that help identify that a user is posting the tweet from a location. 
\end{enumerate}

\begin{enumerate}[(g)]
    \item Lemmatization and PoS Tagging. All publications are pre--checked and cleaned before lemmatization for better results. We used the Freeling tool, which is a library providing language analysis functionalities (morphological analysis, named entity detection, PoS--tagging, parsing, etc.) for a variety of languages \citep{padro12}.The results are shown in Table \ref{tbl6}.
\end{enumerate}

\begin{table*}[t]
\caption{Text output after lemmatization and PoS tag with Freeling.}
\label{tbl6}
\centering
\begin{tabular}{p{5.5cm}p{5.5cm}p{4cm}}
\toprule
Original Text & Lemmatized text & Part of Speech \tabularnewline
\midrule
I'm at Mallplaza Bellavista - @mallplazaperu in Bellavista, Callao https://t.co/brtyxSe8CY & estar en centro comercial bellavista en centro comercial en bellavista\_callao & estar/VMI en/SP\newline centro/NC  comercial/AQ\newline bellavista/NP en/ SP\newline centro/NC comercial/AQ\newline en/SP bellavista\_callao/NP
\tabularnewline
\hline
work breakfast! \#friends \#meeting en Universidad Jorge Tadeo Lozano https://t.co/sNYJhxG6cw & trabajo desayuno amigo reunión en universidad jorge\_tadeo\_lozano & trabajo/NC desayuno/NC\newline amigo/AQ reunión/NC\newline en/SP universidad/NC\newline jorge\_tadeo\_lozano/NP \tabularnewline
\hline
Un dia cualquiera en Cevicheria Karloncho Oficia https://t.co/f9kdEEwdMx & día cualquiera en restaurante pescado oficia & día/NC cualquiera/PI\newline en/SP restaurante/NC\newline pescado/NC oficiar/VMI
\tabularnewline
\hline
Terminando de cantar la Santa Misa dominical \text{\#santamisa} \text{\#musicacatolica} \text{\#singer} \text{\#catholic} \text{\#church} en Capilla \text{Jesus} Hostia \newline \text{https://t.co/pLpvHwRNYh} & terminar cantar santo misa dominical cantante iglesia en capilla jesús\_hostia & terminar/VMG cantar/VMI\newline santo/NC misa/NC\newline dominical/AQ cantante/NC\newline iglesia/NC en/SP capilla/NC\newline jesús\_hostia/NP \tabularnewline
\hline
He venido a que \text{mami} me atiborre de comidaaaaaa (@ Residencial Parque Central in Lima) https://t.co/dCCBbEgvZm & venir mamá atiborrar en residencial parque central & venir/VMP mamá/NC\newline atiborrar/VMS en/SP\newline residencial/NC\newline parque/NC central/AQ \tabularnewline
\bottomrule
\end{tabular}
\end{table*}

\begin{figure}
	\centering
		\includegraphics[width=1\linewidth, scale=1]{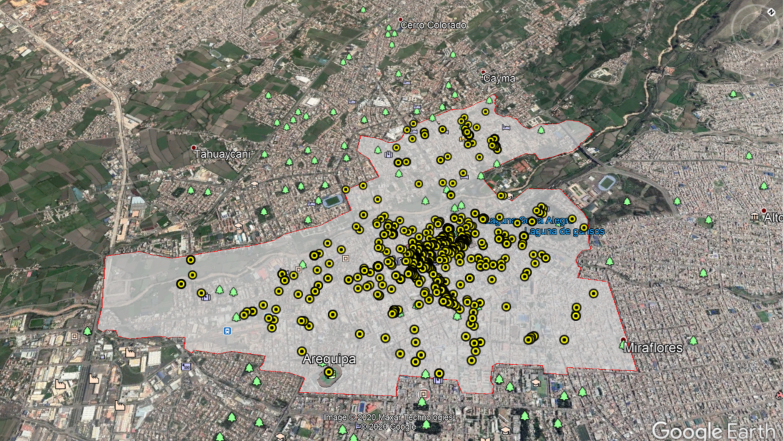}
	\caption{Data located in the historical center.}
	\label{FIG:5}
\end{figure}

\begin{figure}
	\centering
		\includegraphics[width=0.8\linewidth, scale=1]{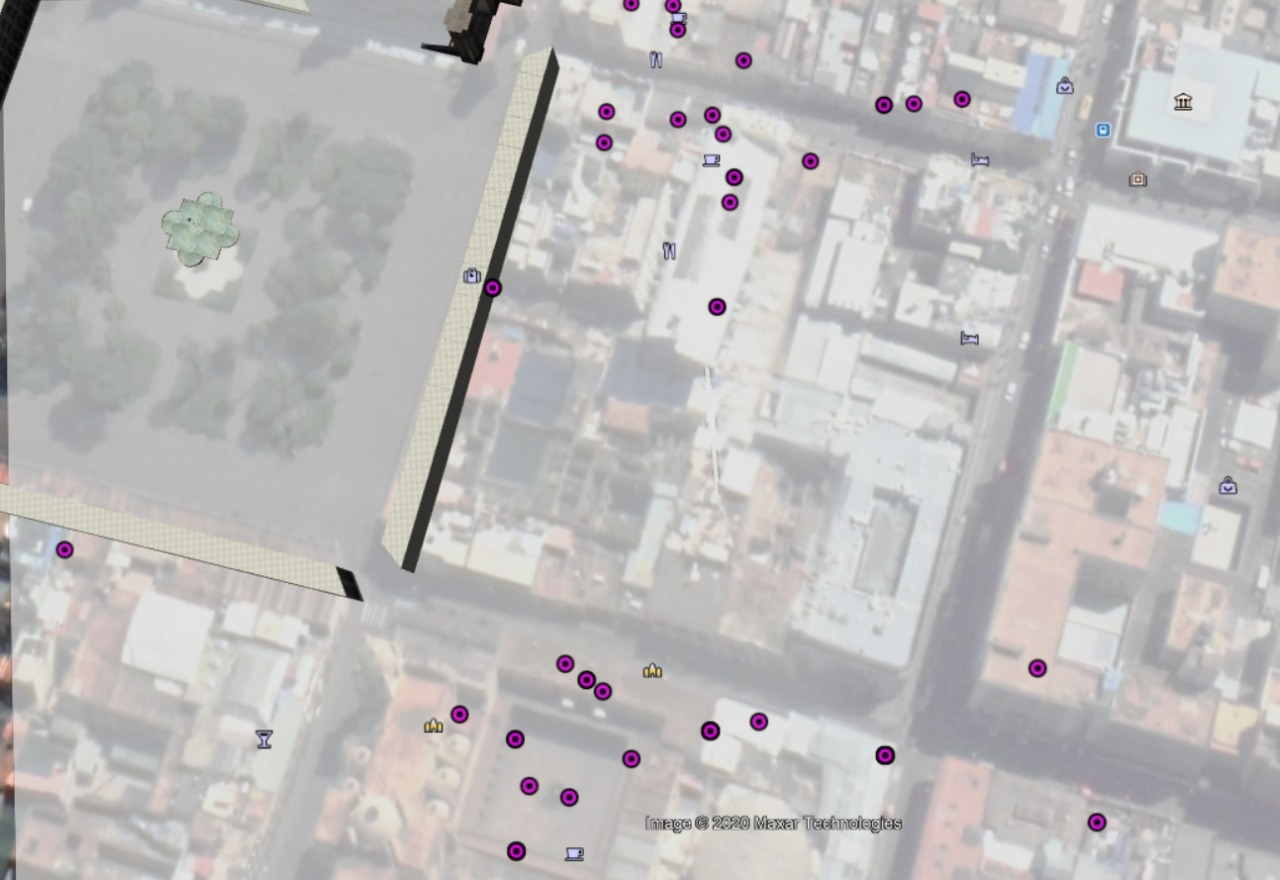}
	\caption{Detailed example from the second cleaning process.}
	\label{FIG:7}
\end{figure}

After pre--processing, the tweets belonging to the application data are selected according to their geolocation (South America). Each tweet positioned in the radius of the city of Arequipa goes through two conditions: the first condition selects the tweets that are located within the polygon that represents the historical center, and the second condition selects the tweets that are located in a polygon that represents a block. To perform the first filtering, the \textit{shapefile} containing the representation of the historical center is imported into the PostgreSQL database and the tweets geo--positioned in Arequipa (ID, latitude, and longitude) are loaded into a \textit{tweet\_gps} table. To select the tweets that are located in the polygon, the coordinates are transformed into \textit{geometry} type data with the function \textit{st\_geomfromtext}. For instance, the SQL query used returned a total of 2343 tweets distributed in the plane as shown in Fig. \ref{FIG:5}.  For the second filter, the polygons of the blocks are stored in the PostgreSQL database as \textit{geometry} type data. Tweets that are located in the middle of the street or outside the historic center are removed, providing 924 tweets (see in Fig. \ref{FIG:7} a zoomed area).


%

\subsubsection{Feature Extraction}

Using only one feature extraction method does not guarantee the best results according to \citep{tessore}. As a consequence, this research explores the use of TF--IDF values associated with a multidimensional vector, the \textit{n--grams} (unigrams, bigrams, and trigrams), which have shown to work well in the classification of documents \citep{anzovino}, and \textit{Bag--of--PoS} that works similarly to the \textit{n--grams}, but based on the sequence patterns of the Part--of--Speech labels. 

For example, according to \citep{sakaki}, the \textit{``Verb-Preposition''} pairs are commonly followed by the name of the place where a user is located, which allows it to be used as an indicator to identify whether the user is talking about an establishment or location. Once the data is numerical, the categorization of tweets becomes a standard machine learning classification problem.

\subsubsection{Classification}

The Multinomial Naïve-Bayes Classifier (MNB) is one of the most popular algorithms in social media data and text categorization \citep{Kateb,anzovino}. It is based on the evaluation of the probabilities of each class. Each tweet in the corpus has associated vectors with the extracted characteristics and their respective label of the land use type; therefore, the classifier is trained with each one of the characteristics, evaluating them individually, and later different combinations are examined.

Unlike applying the classifier to the corpus data which was filtered in the labeling process ensuring that they all refer to one location, when using the classifier with the application data, there is a need to previously identify which tweets refer to the user being in one location and which refer to other topics. For that purpose, with the tagged and untagged corpus data, we selected the PoS sequences with Bag--Of--PoS and identified the \textit{i--most frequent} sequences that represent each data set, as it is done in \citep{Koto}. Depending on the presence or not of the sequences of each class in the text ([it is / it is not] in a location), the tweet is classified by the MNB approach.

\subsection{Results}
\subsubsection{Land use categorization}

Class imbalance and expected results are some of the considerations taken into account in the selection of metrics to measure the performance of a classifier. There is no one metric that measures well the performance of a classifier in every scenario. In this research, the classifier is of the multi--class type (many possible results that are mutually exclusive) \citep{Sokolova} , so it is also important to calculate the metrics for each class \begin{math}c_{1}...c_{n}\end{math} individually, and then calculate the classifier metrics generally, as proposed in \citep{tellez}:  accuracy, precision, recall, and F1--score. Since the classes are unbalanced (some classes have more data, so they are more likely to appear than others) the prioritized indicator is the F1--score \citep{anzovino}. Results for each model are shown in Table \ref{tbl7}.

\begin{table}[t]
\caption{Metrics with different combinations of characteristics.}
\label{tbl7}
\centering
\begin{tabular}{p{3.8cm}p{1.2cm}p{1.2cm}p{0.8cm}p{1.4cm}}
\toprule
Feature & Accuracy & Precision & Recall & F1-score \tabularnewline
\midrule
TF-IDF & 0.830 & 0.911 & 0.702 & 0.744 \tabularnewline
TF-IDF / lemma & 0.836 & 0.907 & 0.730 & 0.772 \tabularnewline
\hline
Unigram  & 0.884 & 0.896 & 0.810 & 0.838 \tabularnewline
Unigram / lemma & 0.884 & 0.886 & 0.832 & 0.851 \tabularnewline
\hline
Bigram  & 0.837 & 0.887 & 0.770 & 0.805 \tabularnewline
Bigram/ lemma & 0.843 & 0.868 & 0.784 & 0.813 \tabularnewline
\hline
Trigram  & 0.615 & 0.843 & 0.492 & 0.559 \tabularnewline
Trigram/ lemma & 0.651 & 0.855 & 0.551 & 0.624 \tabularnewline
\hline
N-gram (1,2,3) & 0.894 & 0.899 & 0.844 & 0.863 \tabularnewline
N-gram (1,2,3) / lemma & \textbf{0.904} & \textbf{0.900} & \textbf{0.870} & \textbf{0.880} \tabularnewline
\bottomrule
\end{tabular}
\end{table}

The corpus is used with two variations: one, with the text in its lemmatized form and the other with the original text, without lemmatization.  The values achieved with these two variations are very close to each other, but the use of the lemmatized text always provides slightly better results (see Table \ref{tbl7}, where two rows for each feature is shown, with and without lemma). Furthermore, comparing the results of the models it is clear that by using all n--grams (unigram + bigram + trigram) together the highest accuracy is achieved. Therefore, this model is chosen for the classification of application data resulting in tweets labeled to be positioned in a Land Use Map (a sample of the results achieved with real data is shown in Table \ref{tbl8}).

\begin{table*}[h]
\caption{Classification results with real data.}
\label{tbl8}
\centering
\begin{tabular}{p{8.5cm}p{2cm}p{2cm}p{2.2cm}}
\toprule
Original Text & Latitude & Length & Label \tabularnewline
\midrule
Buenas noches Arequipa!!  (@ Zig Zag in Arequipa) https://t.co/SY97QrTo9e https://t.co/v4UNcnljT0 & -16.39525055 & -71.53541831 & Commercial \tabularnewline \hline
Amando conocer este pais [?] en Plaza de Armas de Arequipa https://t.co/rBZ4Dmw0iI & -16.39869651 & -71.53693914 & Unbuilt land \tabularnewline \hline
\text{\#Arequipa} \text{\#Arte} \text{\#Concierto} de bienvenida en la inauguracion de la \text{\#Exposicion} \text{\#Raices} @ Centro De Las Artes De La Ucsp \text{https://t.co/f76n4QdHks} & -16.3998186 & -71.5393672 & Institutional \tabularnewline \hline
Tarde de películas en casa \text{\#amor} \text{\#dulcehogar} \text{\#movie} time & -16.399428 & -71.539881 & Residential \tabularnewline \hline
Saliendo de la oficina \text{\#work} en Galeria San Jose \text{https://t.co/BvkDpu2zi6} & -16.3988135 & -71.5318563 & Industrial \tabularnewline \hline
\text{\#Plus135}: Que son las condiciones objetivas de punibilidad? \text{https://t.co/64doI8zoLI} & -16.3902511 & -71.5360128 & Non--classified \tabularnewline \hline
\bottomrule
\end{tabular}
\end{table*}

Out of the 924 tweets used by the classifier, 327 were classified in the commercial category, 248 in the institutional category, 35 in residential, 14 in unbuilt land, and 8 in industrial. There were 292 tweets left that did not reach the cut--off point established for the classifier, so they were not labeled (non--classified) assuming that the text does not refer to a location.

\subsubsection{Results visualization}

The tweets were also presented on the cadastre map of the historic city center. In this way, it is possible to compare the land uses identified by the classifier against the land use category registered in the cadastre by the local municipality. 

\begin{figure*}
	\centering
		\includegraphics[width=1\linewidth, scale=1]{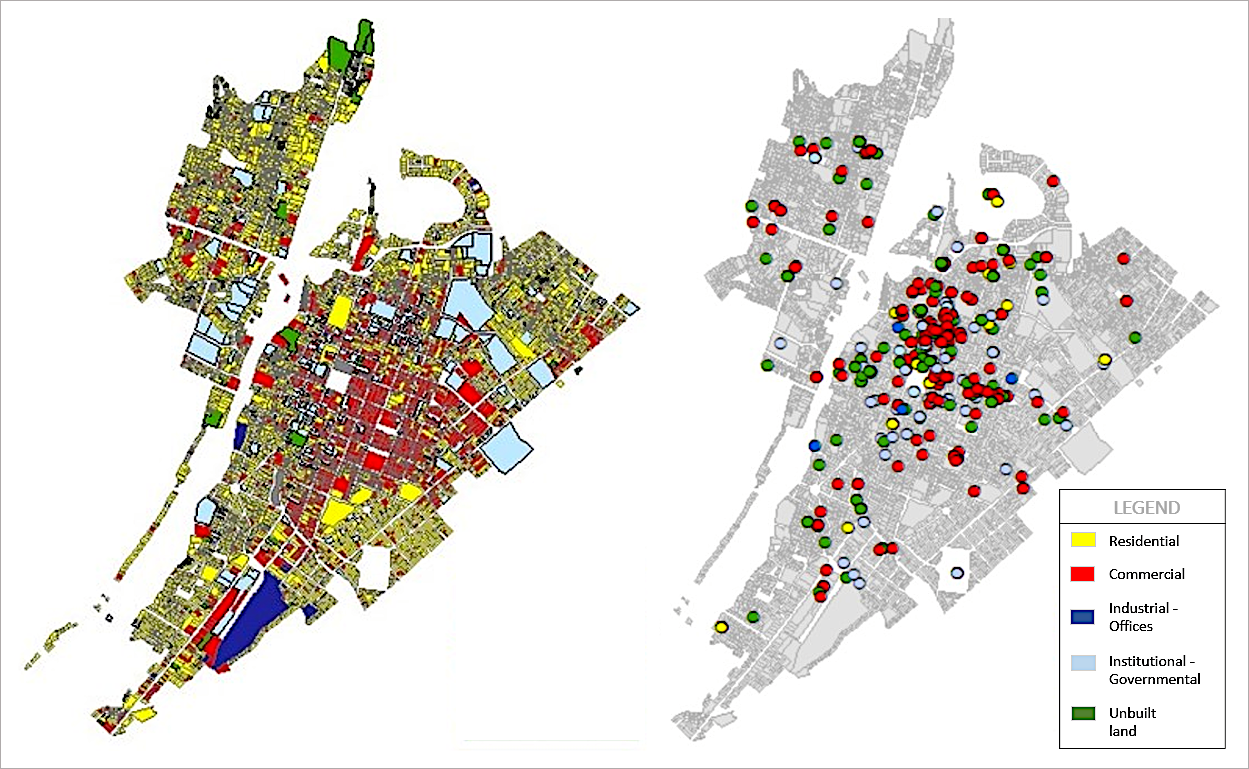}
	\caption{The image on the left shows the land use map of the historical center (cadastre); the image on the right shows the tweets labeled on the map by our approach.}
	\label{FIG:11}
\end{figure*}

The map of the historic center is shown in Fig. \ref{FIG:11} (on the left) with the classification of land uses according to the city's cadastre. On the map, each category of land use was associated with a color. Thus, the use of residential, commercial, industrial--office, institutional--governmental, and unbuilt land have the colors yellow, red, blue, light blue, and green, respectively. The second map in  Fig. \ref{FIG:11} (on the right) shows the tweets labeled with the categories of land use. According to the information shown on the map, a large number of land uses corresponding to commercial, industrial, unbuilt land, and institutional--governmental were identified. However, a small number of tweets with the residential label is observed although according to the cadastre a significant percentage of land uses in the historic center corresponds to residential. Therefore, there is no correspondence on the residential category between the cadastre and the automatic classification of land use. 

\begin{figure}
	\centering
		\includegraphics[width=0.5\linewidth, scale=1]{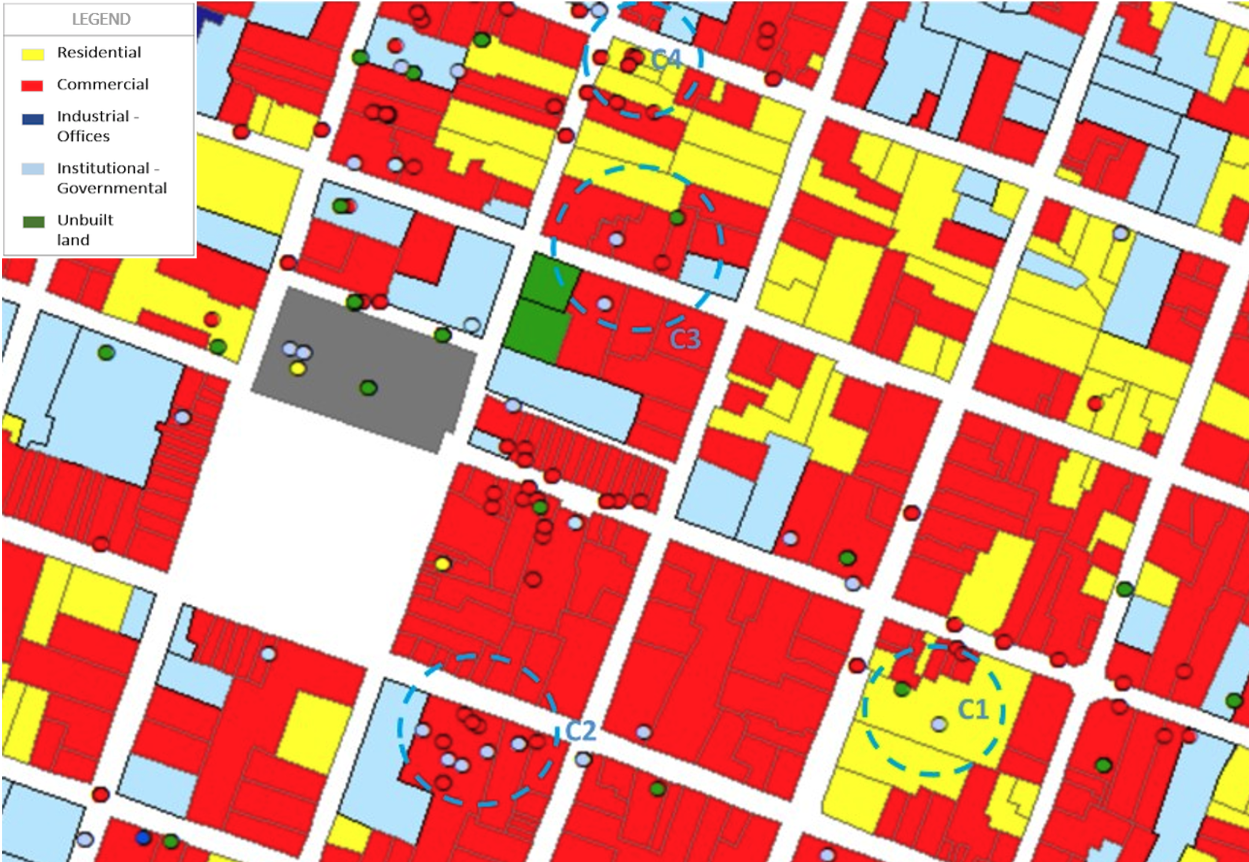}
	\caption{Example of inconsistency between the categorization of the cadastre and the result obtained by the classifier.}
	\label{FIG:12}
\end{figure}

\begin{figure}
	\centering
		\includegraphics[width=0.5\linewidth, scale=1]{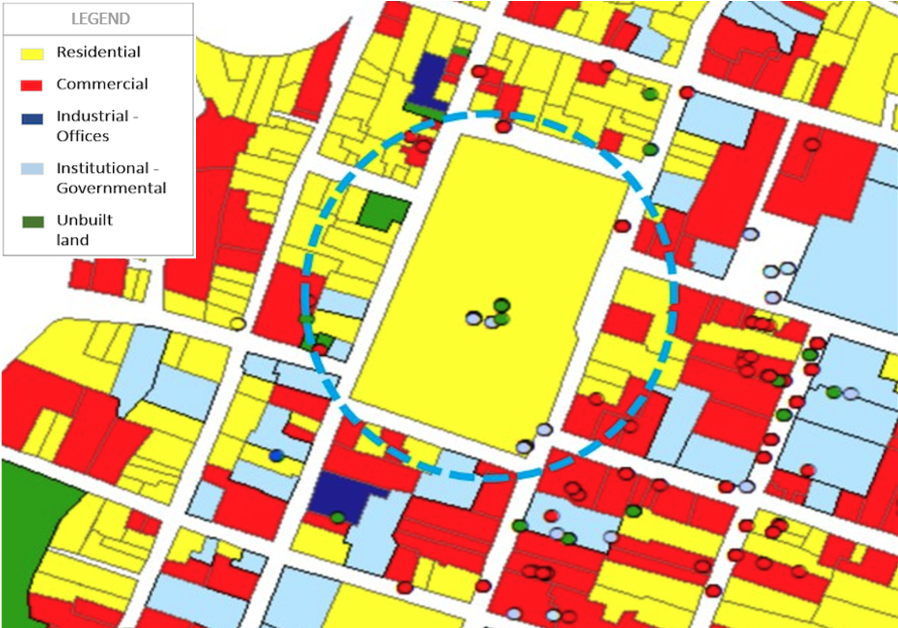}
	\caption{The highlighted area corresponds to the convent of Santa Catalina (in the cadastre it is wrongly classified as residential land use).}
	\label{FIG:13}
\end{figure}

In order to deepen the analysis of the inconsistencies revealed between the two methods, Figs. \ref{FIG:12} and \ref{FIG:13} clearly show interesting differences. Color scale for land use is the same for both, as depicted in the legend. Fig. \ref{FIG:12} illustrates areas (light blue circle) labeled as residential in the cadastre, but identified as vacant land or institutional. In the other hand, Figs. \ref{FIG:13} shows a block (yellow) in which is located the Santa Catalina Convent, and it is cataloged as residential in the cadastre. However, tweets indicate the potential use as unbuilt or institutional--government land (religious is a subcategory within the last one). Multiple cases of inconsistency were detected when comparing cadastre and classifier outcome, many of which were later verified by a site visit. In general, the results provided by our approach were much more accurate than those registered in the cadastre.

\section{Conclusions}

Social networks provide valuable data on urban dynamics, offering new opportunities for research in the field. In this study, we used Twitter data to analyze land use in the historical center of the city of Arequipa by capturing tweets from the area with the text of the publication, time, date, and coordinates.

This research proposes a complete methodology of NLP for the analysis of tweet texts and coordinates, together with the Naïve Bayes Multinomial algorithm for the classification of spaces within land use categories. The evaluation of the model shows that the approach provides excellent results, with accuracy of about 90\%, and F1--score of about 88\%. The information of the area obtained from the project ``Height for Culture'' is used as a basis for the interpretation of the results, verifying that, as expected due to the knowledge of the area, a large percentage of the properties belong to the commercial category because these are in the historical city center with high presence of tourism, followed by the buildings of the institutional--cultural category due to the historical character of the area. However, the methodology detected that many residential spaces registered in the cadastre have currently other activities or uses.

We conclude that the Twitter data provides useful information to identify land uses in the geographic area where it is captured. Tweets are gathered in a simple and inexpensive way and provide information that can be used as an additional method by urban planning professionals and organizations interested in that area.The advantage of this model over the traditional ones resides in its dynamism since it uses data that is constantly updated by the users and allows reflecting the inconsistencies that exist in the maps generated by the cadastre due to the constant change of the environment. Also, the methodology is easily transferred to other geographical areas by means of the use of specific dictionaries to the region under study. This knowledge might be used as a recommendation system for short--term supervision or updating of the cadastre.

Finally, this method depends on the amount of geo--loca\-ted data available at the time of classification, so the capture of publications from other social networks should be considered for future work to maximize the effectiveness and usefulness of the results. 

\section{Acknowledgements}
This research has been funded by the Universidad Nacional de San Agustín de Arequipa with contract IBA 0021-2017-UNSA, by the Junta de Andalucia under grant nº 1155595, and by the Ministry of Science and Innovation under grant PID2020-117759GB-I00.

\bibliographystyle{unsrtnat}    
\bibliography{land_use_identification}

\end{document}